\documentclass[aps,pre,twocolumn,showpacs,superscriptaddress,groupedaddress]{revtex4}
\usepackage{color,url,ulem,bm,amsmath,times,graphicx,epsfig}
\usepackage{subfigure}

\newcommand{\bmx}{\ensuremath{\bm{x}}}
\newcommand{\bmq}{\ensuremath{\bm{q}}}

\newcommand{\bmu}{\ensuremath{\bm{u}}}

\newcommand{\myeq}[1]{Eq.\ (\ref{#1})}
\newcommand{\myeqs}[2]{Eqs.\ (\ref{#1}) and (\ref{#2})}

\newcommand{\myfig}[1]{Fig.\ \ref{#1}}

\newcommand{\Realpha}{\ensuremath{{\text{Re}[\alpha]}}}

\newcommand{\Deff}{D^{\text{eff}}}

\newcommand{\hide}[1]{{}}
\newcommand{\strike}[1]{{}}
\begin{document}

\title{Shear flow controlled mode selection in a nonlinear autocatalytic medium}

\author{S.G. Ayodele}
\affiliation{
Max-Planck-Institut f\"ur Dynamik und Selbstorganisation, Am Fassberg 17, 37077 G\"{o}ttingen, Germany.}
\author{D. Raabe}
\affiliation{Max-Planck Institut f\"ur, Eisenforschung, 
Max-Planck Stra{\ss}e 1, 40237, D\"usseldorf, Germany.}
\author{F. Varnik}
\thanks{Corresponding author: fathollah.varnik@rub.de}
\affiliation{Interdisciplinary Center for Advanced Materials Simulation (ICAMS),
Ruhr-Universit\"at-Bochum, Universit\"atsstra{\ss}e 150,  44789 Bochum, Germany.}

\begin{abstract}
The effect of shear flow on mode selection and the length scale of patterns formed in a nonlinear auto-catalytic reaction-diffusion model is investigated. We predict analytically the existence of transverse and longitudinal modes. The type of the selected mode strongly depends on the difference in the flow rates of the participating species, quantified by the differential flow parameter. New spatial structures are obtained by varying the length scale of individual modes and superposing them via the differential flow parameter. Our predictions are in line with numerical results obtained from lattice Boltzmann simulations.
\end{abstract}


\maketitle
\section{INTRODUCTION}
\label{sec:Intro}
Reaction-diffusion models of excitable media are known to give rise to spatially and/or temporally varying patterns under certain conditions. The formation of these patterns in stationary media can be explained based on a mechanism first proposed by Turing~\cite{Turing1952}. However, in moving excitable media, the  mechanism of pattern formation, length scale of the patterns formed and  the type of patterns selected can be significantly different from the stationary case. Under a constant uniform flow for instance, differences in the flow rates of the participating reacting species can spatially disengage the  species leading to the well known differential flow instability (DIFI) mechanisms~\cite{Andrésen1999,Kaern1999,Kaern2000,Kaern2002,Stucchi2013}. Patterns formed from this mechanism are known to be entirely different from those resulting from the Turing mechanism occurring in stationary media. In a linear shear flow, the effect of the distribution of fluid velocities has been shown to give rise to 
another mechanism of pattern formation which does not require  differences in flow rate of the reacting species or the fulfillment of Turing condition~\cite{Evans1980,Spiegel1984,Vasquez2004,Kuptsov2005}. The assumption of a reacting fluid comoving with the average flow velocity is not sufficient to explain these effects. A variety of different approaches have been proposed to study this problem. In Ref.~\cite{Vasquez2008}, for example, a system of two coupled layers, moving at a constant velocity with respect to one another, is considered. This approach reproduces some of the key features of pattern formation in a shear flow including situations where the reactants move at different flow rates but with the same diffusion rate~\cite{Stucchi2013}. However, these studies are performed for a piecewise defined velocity and differential flow rates  for a 1D model and do not include the combined 
effects of shear, differential advection and differential diffusivity of the reacting species on pattern selection. The roles played by both differential advection and diffusion of the reacting species on pattern selection can often not be separated. As we have recently shown \cite{Ayodele2011,Ayodele2013}, such a situation may be relevant when modeling, e.g.,  growth kinetics and spatial distribution of vegetation~\cite{Stelt2012,Dagbovie2014}. In this case, one of the reacting species (water) undergoes convection while the other component (vegetation) is transported via diffusion only. Other examples here are interaction of a flowing chemical species with another chemical species bound to a catalytic surface as in packed bed reactors~\cite{Yakhnin1996,Menzinger2004} or cell polarization in developmental biology~\cite{Goehring2011}. 

In this work, we address this issue both analytically as well as via 2D numerical simulation of the nonlinear equations. Via a linear stability analysis of the advection-diffusion-reaction equations, we first determine the dispersion relation for the growth rate of perturbations for an arbitrary differential flow parameter. A simple scaling ansatz then allows to also include the effect of shear rate in the characteristic length scale of the patterns. We find that, by tuning the differential flow parameter, it is possible to select between transverse, longitudinal or mixed modes. In the latter case, the interaction of the selected modes can give rise to interesting novel structures, where the underlying length scale is tuned by shear rate.

\section{The Model}
\label{sec:Model}
As a prototypical example, we choose the two species Gray-Scott model with an advection term to illustrate the effect of shear flow on pattern formation. The most general situation, as observed in aqueous solutions of reactants, is the coupling of the nonlinear reaction to the ambient fluid flow via the density or viscosity of the fluid~\cite{Vasquez1993,Hejazi2010}. Such situations lead to hydrodynamic instabilities and fluxes in the horizontal layer of the reacting fluid~\cite{Bewersdorff1987}. Rather than considering this case, we focus here on the simple situation of the so called 'passive advection', where the flow field is imposed externally and decoupled from the reacting species. The resulting  advection-diffusion-reaction (ADR) equation describing the transport of the two species A and B can then be written as~\cite{Ayodele2011} 
\begin{small}
 \begin{eqnarray}
\tau_A\frac{\partial A}{\partial t}+\tau_A\bmu\cdot\nabla A &=& \left( 1-A\right) -B^{2}A+\tau_A D_{A}\nabla^{2}A, \label{eq:adr1A}\\
\tau_B\frac{\partial B}{\partial t}+\tau_B\delta \bmu\cdot \nabla  B &=& - B +\eta B^{2}A+\tau_B D_{B}\nabla^{2}B,
\label{eq:adr1B}
\end{eqnarray}
\end{small}
where $A=A(x,y,t)$ and $B=B(x,y,t)$ are the dimensionless concentrations of the interacting species. The time scales $\tau_A$ and $\tau_B$ are the characteristic time  for the addition and removal of the species A and B respectively. They are related to the rate constants $k_f$ and $k_2$ as, $\tau_A =1/k_f$ and $\tau_B=1/(k_f+k_2)$~\cite{Ayodele2011}. The  parameter $\eta$ sets the activation of the interacting species and is related to the reaction rate constants $k_1$ as $\eta =  A_{0}(k_{1}k_{f})^{1/2}/(k_{f}+k_{2})$~\cite{Ayodele2011}. $D_A$ and $D_B$ are the diffusion coefficient of the two species and $\bmu(x,y)$ is the local flow velocity. The parameter $\delta$, designated here as the differential flow parameter, is the ratio of the flow rates of the two interacting species. For simplicity, we consider here the flow between two parallel walls driven either by gravity (Poiseuille) or by the relative motion of the walls with respect to one another (Couette). For the Poiseuille flow, the velocity 
profile reads $\bmu(y) = u(
y) \hat{\bmx} = (6u_\text{avg}/L^2_y) y (L_y-y) \hat{\bmx}$. Here, $\hat{\bmx}$ is the unit vector along the $x$ direction, $u_\text{avg}$ is the average fluid velocity and $L_y$ denotes the channel width. For the planar Couette flow,  $\bmu(y) = \dot{\gamma} y \hat{\bmx} $, with a constant shear  rate, $\dot{\gamma}$.

In the absence of advection, the system in \myeqs{eq:adr1A}{eq:adr1B} admits three spatially homogeneous steady state solutions~\cite{Ayodele2011}. The first solution is the trivial homogeneous state $B_{e}=0$, $A_{e} = 1$ and exists for all system parameters. The other two solutions exist provided that the parameter $\eta > 2$. They are given by
\begin{equation}
 A^{\pm}_{e} = \dfrac{\eta\pm\sqrt{\eta^{2}-4}}{2\eta},\qquad \text{and}\qquad B^{\pm}_{e} = \dfrac{\eta\mp\sqrt{\eta^{2}-4}}{2}.
\label{eq:adr2}
\end{equation}
These solutions are destabilized by Turing and/or Hopf instability, leading to spatially and/or temporally varying structures. The conditions for Turing and Hopf instabilities in the system are $D_A/D_B>(\tau_B/\tau_A)\eta B^-$ and  $\tau_A/\tau_B>\eta B^-$, respectively \cite{Ayodele2013}.

\section{Results}
\label{sec:Results}
\subsection{Numerical simulation}
A first glance of flow effects on the resulting patterns is obtained by solving \myeqs{eq:adr1A}{eq:adr1B} numerically in a rectangular domain with an imposed Poiseuille flow along the $x$-direction. The numerical solution is performed with the lattice Boltzmann method~\cite{succi2001}. Via a multiscale expansion analysis, we have shown in a previous work that, adding appropriate reaction terms to the standard lattice Boltzmann method allows to recover \myeqs{eq:adr1A}{eq:adr1B} \cite{Ayodele2013}. 
The channel dimension is $100 \times 30$ lattice units. Periodic boundary condition is applied along the flow direction. At the walls of the channel, a no-flux boundary condition is imposed for the concentration field, while the no-slip condition is applied to the fluid velocity. The initial condition consists of a random perturbation to the steady state solution \myeq{eq:adr2} at the channel inlet. Note that this is 
different from a fixed profile as used in~\cite{Kuptsov2005}. The diffusion coefficients of the two species are chosen such that the ratio $D_A/D_B$ satisfies the condition for Turing instability~\cite{Ayodele2011}.

\begin{figure}
\includegraphics[scale=0.46]{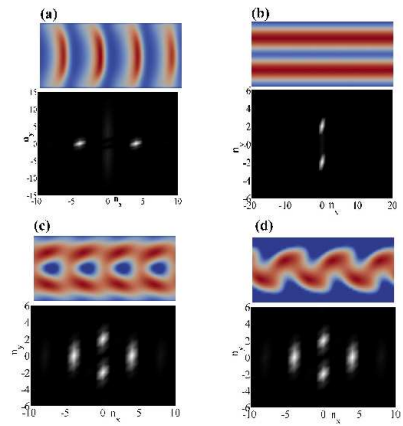}
\caption{(Color online) Patterns obtained from numerical simulations of the model at different differential flow parameters, (a) $\delta$ = 0,  (b) $\delta$ = 1, (c) $\delta$ = 0.7 [excitation of both the modes in (a) and (b)]. (d) $\delta$ = 0.7[consisting of the mode in (a) and a mode similar to (b) but containing a single stripe]. The system parameters  are chosen as $\eta = 2.1988$,$D_A/D_B=8.0$,$\tau_A/\tau_B = 2.744$ and $\dot\gamma =0.001$ and are the same in each panel except in (d) where  $D_A/D_B=4.0$.}
\label{fig:modes}
\end{figure}

Results obtained from these simulations are depicted in \myfig{fig:modes}. We observe longitudinal and transverse modes and superposition thereof depending on the value of the differential flow parameter. Note that these patterns are stable and steady over  time compared to the time for the slowly diffusing species to sample the width of the channel (i.e. for $t \sim L^2_y/D_B$). Here, ``transverse'' and ``longitudinal'' refer to the orientation of the wavefront with respect to the flow direction. If the wavefront is perpendicular (parallel) to the flow direction then the mode is called transverse (longitudinal). Above a critical shear rate, advecting one of the species, while keeping the other immobile ($\delta =0$) leads to a transverse mode with wave vector $\bmq=(q_x,0)$ (\myfig{fig:modes}(a)). Further increase in the shear rate does not change the type of the mode selected, it however leads to the change in wavelength of the pattern. Advecting the two 
species at the same flow rate ($\delta =1$) leads to longitudinal modes with wave vector $\bmq=(0, q_y)$ (\myfig{fig:modes}(b)). In this case, both the type of the selected mode and its wavelength are independent of shear rate. For $1<\delta<0$, both the longitudinal and transverse modes are exited, giving rise to interesting spatial structures as exemplified in Figs.~\ref{fig:modes}(c) and (d). Figure \ref{fig:modes}(c), for instance, is obtained from the interaction of the four stripe transverse mode in (a) with the two stripe longitudinal mode in (b), while the structure in (d) is obtained from the interaction of the four stripe transverse mode in (a) with a single-stripe longitudinal mode.

\subsection{Shear dispersion effects}
The structures observed above in \myfig{fig:modes} are due to the inhomogeneity of the flow~\cite{Ayodele2013}. To rationalize the dispersion effect introduced by the inhomogeneous flow field, we note that the characteristic time scales for the addition and removal of the species A and B are $\tau_A$ and $\tau_B$. Within these time scales, species diffuse over length scales $l_A = (2D_A\tau_A)^{1/2}$ and $l_B = (2D_B\tau_B)^{1/2}$. We have shown in previous work that these length scales determine the characteristic length of the resulting patterns in the absence of flow \cite{Ayodele2011}. Here we go one step further and show that modification of these length scales by the flow provides a simple way of tuning the characteristic length of the pattern along the flow direction and incorporating the effect of shear flow. For this purpose, we consider thermal diffusion of a particle of species A across streamlines of different velocities. Within a time scale of the order of $\tau_A$, the particle diffuses a 
distance of the order of $\Delta y_0 = \sqrt{2 D_
A \
\tau_A}$ and experiences a change in the flow velocity of the order of $\Delta u =  \dot\gamma \Delta y_0$. Its displacement along the direction of flow thus contains both a contribution due to thermal motion, $\Delta x_0$, and from the change of flow velocity, $\Delta x^{\text{flow}}_A=\tau_A\Delta u_A=\tau_A\dot\gamma\Delta y_0$. One can, therefore, write
\begin{eqnarray}
(l^{\text{eff}}_{x,A})^2 &=& \langle(\Delta x_{0}+\Delta x^{\text{flow}})^{2}\rangle \nonumber\\
&=&  2 D_A \tau_A + (\tau_A \dot{\gamma})^2 2 D_A \tau_A  \equiv 2 \Deff_{x,A} \tau_A \label{eq:DeffA}\\
l^{\text{eff}}_{x,A}  &=& l_A  \sqrt{1 + (\tau_A \dot{\gamma})^2},
\label{eq:lA}
\end{eqnarray}
where we used the isotropy of thermal diffusion and the fact that the displacements due to the flow and diffusion are not correlated $\langle\Delta x_{0} \Delta x^{\text{flow}}\rangle=0$. The last equality in \myeq{eq:DeffA} defines the flow-enhanced effective diffusion coefficient along the flow direction $\Deff_{x,A}  = D_A  \big[ 1 + (\tau_A \dot{\gamma})^2 \big]$. Similarly, and noting that $\Delta u_B=\dot\gamma\Delta y_0 \delta$, one obtains for the species B, 
\begin{equation}
l^{\text{eff}}_{x,B} = l_B \sqrt{ 1 + (\tau_B \dot{\gamma}\delta)^2}
\label{eq:lB}
\end{equation}
and $\Deff_{x,B} =  D_B \big[ 1 + (\tau_B \dot{\gamma}\delta)^2 \big]$. It is noteworthy that our results differ from the well-known Taylor dispersion by the fact that effective diffusion in the former continuously increases with time ($D^\text{eff, Taylor}(t)=D[1+(\dot\gamma t)^2]$) \cite{Varnik2007},  while in our case, characteristic time scales $\tau_A$ and $\tau_B$ set a limit to this process.

The validity of \myeq{eq:lA} is examined by choosing $\delta = 0$, and performing  series of simulations for two different flow situations; a planar Couette flow  and a Poiseuille flow between two planar plates. The effective characteristic length scale $l_A$ is obtained from the simulation results by computing the characteristic wavelength $\lambda$ for each pattern. As shown by linear analysis of the ADR equations for a constant uniform flow\cite{Ayodele2013}, the wavelength of the patterns is related to the characteristic length scale  $l_A$ via ($q=2\pi/\lambda$)

\begin{equation}
q^2 =\frac{(l^{\text{eff}}_{x,A}/l_{B})^2 u^2_{\text{avg}} - \eta B_e}{2(l^{\text{eff}}_{x,A})^2}.
\label{eq:lambda_c}
\end{equation}

\begin{figure}
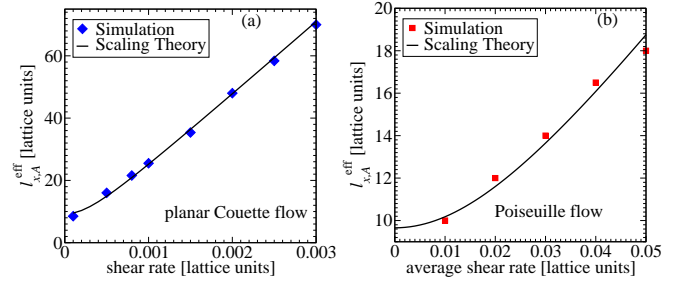

\begin{center}
\includegraphics[scale=0.18]{Fig2a.eps}
\includegraphics[scale=0.18]{Fig2b.eps}
\end{center}
\caption{(Color online) Simulation results on the characteristic pattern length along the flow direction, $l^{\text{eff}}_{x,A}$ (symbols), versus the prediction of the simple scaling theory, \myeq{eq:lA} (solid lines). The simulation parameters are chosen as $\delta =0$,  $l_A/l_B = 4.75$, $\tau_A = 3030$, $\tau_B = 1104$  and $\eta = 2.1988$. The initial configuration in all cases shown in the panel correspond to  random perturbations added to the homogeneous steady state \myeq{eq:adr2}  at the inlet of the channel. Panel (a) shows the result obtained for a simple shear flow (planar Couette geometry), where the shear rate is constant across the channel. Panel (b) corresponds to a parabolic velocity profile between two parallel plates (Poiseuille flow). Note that, in this case, the shear rate varies across the channel. Therefore, we use the average shear rate (defined as the mean flow velocity divided by the plate separation) when evaluating \myeq{eq:lA}. Obviously, this simple scaling result provides a 
good approximation 
to the simulated data.}
\label{fig:lA}
\end{figure}

Investigating the case $\delta = 0$ allows one to focus on the modulation of species A, while species B is immobilized ~($l^{\text{eff}}_{x,B} = l_B$). As seen from \myfig{fig:lA}, the simple scaling theory accounts quite well for the variation of the characteristic length of the pattern with the flow. It is noteworthy that, while shear rate is constant across the channel in the first case, it varies linearly with distance from the channel center in the case of the Poiseuille flow. Here, the shear rate appearing in \myeq{eq:lA} is estimated by taking the ratio of the average flow velocity to the plate separation, $\bar{\dot\gamma}=u_{\text{avg}}/H$. The good agreement between theory and simulation in this case (\myfig{fig:lA}) underlines the robustness of the scaling analysis, leading to \myeq{eq:lA}.

While tuning the magnitude of flow allows to set the length scale of the patterns, as already shown in \myfig{fig:modes}, differential flow parameter, $\delta$, provides a mean to control mode selection. 

In the light of the above results, this observation can be put on a more rational basis. At first order, the flow modulates the diffusion coefficient in the flow ($x$-) direction while in the $y$-direction it remains unchanged, \myeq{eq:DeffA}. Incorporating this effect and non-dimensionalizing time and length in \myeqs{eq:adr1A}{eq:adr1B} with a characteristic time scale $\tau_A$ and length scale $l_A$, one can rewrite these equations as 

\begin{eqnarray}
\frac{\partial A}{\partial t}+u_x \dfrac{\partial A}{\partial x}=\nonumber &=& \left( 1-A\right) -B^{2}A+\nabla^{2}A \nonumber\\
&&+(\dot{\gamma}\tau_A)^{2}\dfrac{\partial^{2} A}{\partial x^{2}},
\label{eq:adr2A}\\
 \frac{1}{\tau}\frac{\partial B}{\partial t}+\frac{\delta}{\tau} u_x \dfrac{\partial B}{\partial x}&=& -B +\eta B^{2}A+\frac{1}{\varepsilon^{2}} \nabla^{2}B\nonumber\\
&&+\dfrac{(\delta\dot{\gamma}\tau_B)^{2}}{\varepsilon^{2}}\dfrac{\partial^{2} B}{\partial x^{2}},
\label{eq:adr2B}
\end{eqnarray}

where $\tau =\tau_A/\tau_B $ and $ \varepsilon= l_A/l_B$.

\subsection{Linear Stability Analysis}

In order to determine the selected mode, we perform a linear stability analysis around the non-trivial homogeneous state \myeq{eq:adr2} and set

\begin{equation}
A = A_{e}+\phi_{A}e^{\alpha \tilde{t}}e^{i(\tilde{q}_xx+\tilde{q}_yy)},\quad
B = B_{e}+\phi_{B}e^{\alpha \tilde{t}}e^{i(\tilde{q}_xx+\tilde{q}_yy)}.
\label{eq:ansatz}
\end{equation}
Here, $\phi_{A}\ll A_{e} $ and $\phi_{B}\ll B_{e}$ are perturbation amplitudes and $\alpha$ is the growth rate of the perturbations. Note that the wavevectors are in dimensionless units such that $\tilde{q}_x=q_x l_A$, we have dropped the tilde symbol in subsequent discussion for clarity. Inserting this ansatz into \myeqs{eq:adr2A}{eq:adr2B}, linearizing and requiring a non-trivial solutions for $\alpha$ leads to a characteristic equation for the growth rate of type 
\begin{equation}
\alpha^{2}+(a + i b)\alpha + (c+id) = 0,
\label{eq:dispersion}
\end{equation}
The coefficients  $a$,  $b$,  $c$ and  $d$ in \myeq{eq:dispersion} are functions of the system parameters and the wave vector components $q_x$ and $q_y$. The coefficients are given as

\begin{eqnarray}
a &=& -q^{2}(1+\tau/\varepsilon^{2}) + (\tau-\eta B^{\pm}_{e})\nonumber \\
  &&-q^{2}_x(\dot{\gamma} \tau_A)^{2}(\delta/\varepsilon^{2}+1),\nonumber \\
\label{eq:dispersion-a}\nonumber\\
b &=&  -q_x u_{\text{avg}}(\delta +1),
\label{eq:dispersion-b}\nonumber\\
c&=& \tau q^{4}/\varepsilon^{2}+q^{2}(\tau\eta B^{\pm}/\varepsilon^{2}-\tau-\delta u_{\text{avg}}^{2})\nonumber\\
&&+\tau(\eta B^{\pm}_{e}-2)+\tau q^{4}_x(\dot{\gamma} \tau_A)^{2}/\varepsilon^{2}\nonumber \\
&&+ \tau q^{2}_x(\dot{\gamma} \tau_A)^{2}(\delta\eta B^{\pm}_{e}/\varepsilon^{2}-1)\nonumber\\
&& + (\tau q^{2}_xq^{2}(\dot{\gamma} \tau_A)^{2}/\varepsilon^{2})(\delta + 1),
\label{eq:dispersion-c}\nonumber \\
d &=& u_{\text{avg}}(q^{3/2}(\tau/\varepsilon^{2}+\delta )+ q(\delta \eta B^{\pm}_{e}-\tau)\nonumber\\
&&+q q^{3}_x(\dot{\gamma} \tau_A)^{2} (\delta+\delta\tau )),  
\label{eq:dispersion-d}
\end{eqnarray}

 where ($q^{2}=q^{2}_{x}+q^{2}_{y}$). 
 
 Evaluating the roots of the polynomial in \myeq{eq:dispersion} and separating the solution into the real and imaginary parts one obtains  
\begin{eqnarray}
\text{Re}[\alpha] &=&\dfrac{a}{2}\pm\ \dfrac{1}{2}\sqrt{\dfrac{ r + (a^{2}-b^{2}-4c) }{2}}, \label{eq:growth_ratea}\\
\text{Im}[\alpha] &=& \dfrac{b}{2}\pm\ \dfrac{1}{2}\sqrt{\dfrac{r-(a^{2}-b^{2}-4c)}{2}},
\label{eq:growth_rate}
\end{eqnarray}
where $r = \sqrt{(a^{2}-b^{2}-4c)^{2}+(2ab-4d)^{2}}.$

 A non-zero imaginary part of $\alpha$ indicates the presence of oscillations, while the real part of $\alpha$ tells us whether an infinitesimal perturbation will decay ($\Realpha <0$) or grow ($\Realpha>0$), the latter being of particular interest for pattern formation. Moreover, the mode that has the largest positive real part of $\alpha$ is most probable to be selected.

\begin{figure}
\begin{center}
\includegraphics[scale=0.32]{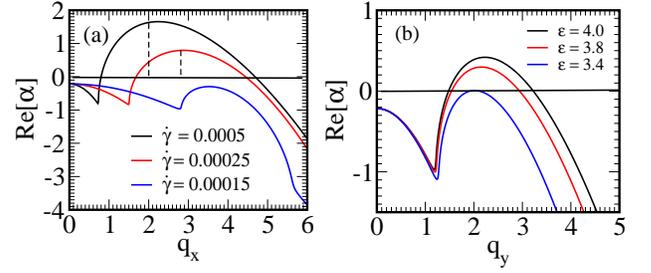}
\end{center}
\caption{(Color online) (a) Effect of shear rate on the growth rate of the $q_x$ mode. The maximum of \Realpha shear rate showing a shift in the length scale of the patterns towards longer wavelength. The parameters are chosen as $q_y=0$, $\delta=0$, $\epsilon=4.75$, $\eta=2.1988$. (b) Effect of $\varepsilon$ on the growth rate of the $q_y$ mode. For $\varepsilon < \eta B_e$ $(\eta B_e = 3.421)$, the $q_y$ mode is damped. The parameters are chosen as $q_x=0$, $\delta=1.0$, $\epsilon=4.75$, $\eta=2.1988$.}
\label{fig:growthrate}
\end{figure}

Figure \ref{fig:growthrate}a shows the effect of shear rate on \Realpha as obtained from an evaluation of \myeq{eq:dispersion}. In line with our scaling analysis of shear-induced diffusion, the wave vector associated with the fastest growing mode (maximum of \Realpha) decreases with increasing shear rate. Another observation from \myfig{fig:growthrate}a is that for a given value of $\varepsilon$, the system becomes unstable upon increasing shear rate  (see $\dot\gamma > 0.00015$). An imposed shear flow thus has two effects: (i) it gives rise to new instabilities and (ii) it influences the length scale of the fastest growing mode.

\begin{figure}
\begin{center}
\includegraphics[scale=0.30]{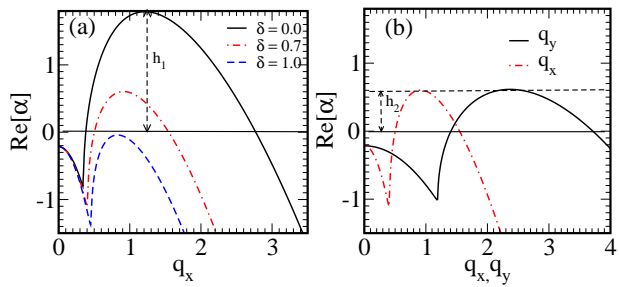}
\end{center}
\caption{(Color online) Plot showing dispersion relations at different differential flow parameters $\delta$ (a) $q_x$ mode (b) comparison of the $q_x$ and $q_y$ mode at $\delta$ = 0.7.}
\label{fig:growthrate2}
\end{figure}

Next we turn to the effect of the differential flow parameter, $\delta$, on the selected mode. For this purpose, we recall that diffusion is enhanced along the flow ($x$-) direction only (cf. \myeqs{eq:lA}{eq:lB}). This property is nicely reflected in the $q$-dependence of the growth rate, obtained from \myeq{eq:dispersion}. Indeed, we find that $\delta$ has no effect on the dependence of the real part of $\alpha$ upon $q_y$, provided that $q_x=0$. In other words, the  excitation and growth of all the modes of the type $\bmq=(0,q_y)$ is independent of the differential flow parameter. As shown in \myfig{fig:growthrate}b, a necessary condition for the excitation and growth ($\Realpha > 0$) of these longitudinal modes is that $\varepsilon > \eta B^{\pm}_{e}$, since all modes with $\varepsilon < \eta B^{\pm}_{e}$ are damped. In order to control mode selection, we first tune $\varepsilon$ in such a way as to obtain a positive growth rate for a longitudinal mode~(see  \myfig{fig:growthrate}b), we then 
modify the growth rate associated with the transverse modes via a variation of $\delta$~(see \myfig{fig:growthrate2}a). As shown in \myfig{fig:growthrate2}b, for $\delta=0.7$, the maximum growth rate for both modes becomes identical. For this set of parameters, both modes coexist (see \myfig{fig:modes}c,d).

\begin{figure}
\includegraphics[scale=0.298]{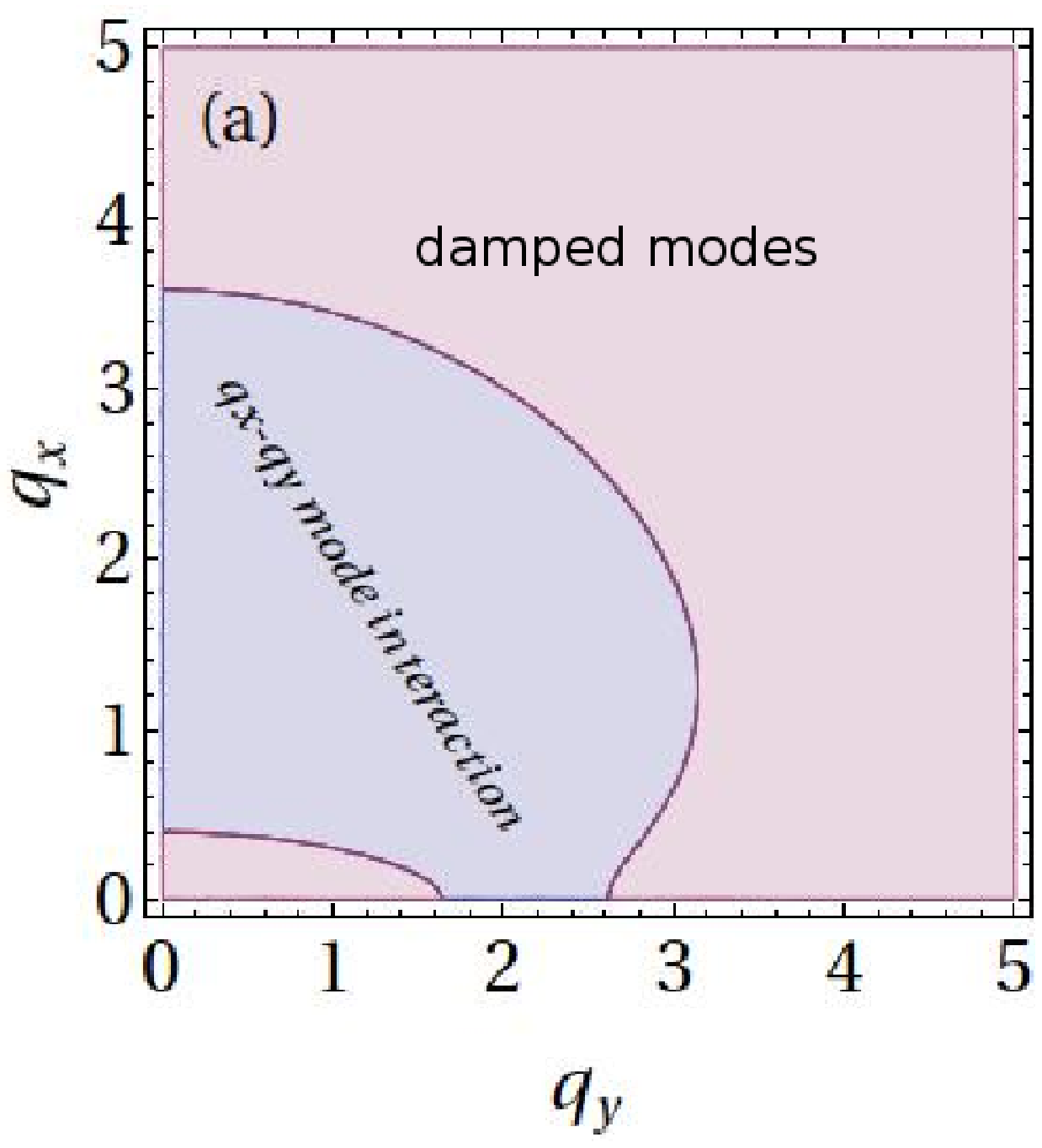}
\includegraphics[scale=0.248]{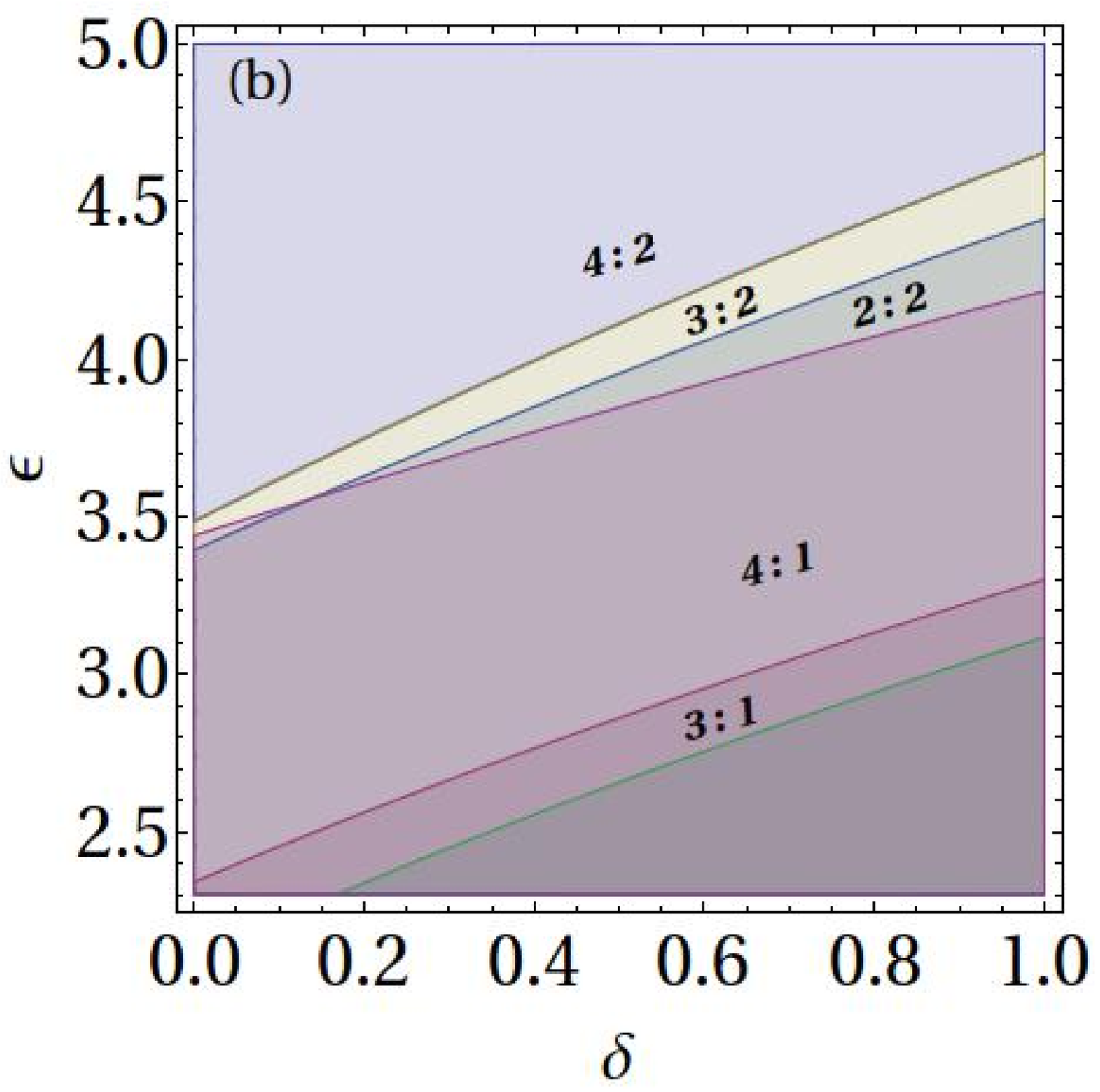}
\caption{(Color online) Stability diagram of excited modes. (a) In the $q_x-q_y$-plane for $\delta= 0.0$ and $\varepsilon = 3.6$. Beyond the grey domain, all the modes are damped. (b) In the $(\varepsilon, \delta)$ parameter space showing the region for the wavenumbers $(n_x:n_y)$ of the modes. The system parameters are chosen as  $\eta = 2.1988$, $\tau = 2.744$ and $\dot\gamma =0.001$.}
\label{fig:modes_2D}
\end{figure}

We numerically compute the stability diagram using the characteristic dispersion relation in \myeq{eq:growth_ratea}. In Figure \ref{fig:modes_2D}a we show a stability diagram of the modes, computed at parameters $\delta =0$ and $\varepsilon = 3.6$. Modes within the grey area are unstable (give rise to patterns) and can coexist or interact, while those outside are damped. To access the stability diagram in $(\varepsilon,\delta)$ plane, we consider modes with wavenumbers that fit into the typical size of our simulation domain~$( q_x= 2\pi n_x/L_x,  q_y = 2\pi n_y/L_y)$ and compute regions in $(\varepsilon,\delta)$ parameter space where the growth rate $\Realpha > 0$. This is shown in Figure\ref{fig:modes_2D}b. In the regions designated by the ratio of the wavenumbers, the corresponding modes can coexist and interact. For instance the modes 4:2 corresponds to the observed pattern in \myfig{fig:modes}c, while the mode 4:1 corresponds to the one in \myfig{fig:modes}d. All other mode interaction can be 
obtained by choosing the corresponding parameters $\delta$ and $\varepsilon$ accordingly. We expect that a wide variety of complex dynamics and spatial resonances can be triggered by the interaction of these steady state modes.

In summary, the effect of shear flow on mode selection and length scale of patterns in nonlinear media is investigated. The  mode selection under shear flow depends on the differential flow parameter of the participating species. Transverse and longitudinal modes are found to be selected depending on the differences in the flow rates of the participating species. Moreover, the effect of a heterogeneous shear on the underlying length scale of the patterns is accounted for via a simple scaling ansatz, which becomes exact at homogeneous shear. The predictions of the model regarding mode selection, length scale and the stability diagram are all in line with numerical simulation results.

\end{document}